\documentstyle[preprint,epsf]{jpsj}
\def\runtitle{Superconductivity of Q2 Dimensional
Tight-Binding Electrons in a Strong Magnetic Field}
\def\runauthor{Mitake {\sc Miyazaki}, Keita {\sc Kishigi} and
 Yasumasa {\sc Hasegawa}}
\title{Superconductivity of Quasi-Two-Dimensional
Tight-Binding Electrons in a Strong Magnetic Field}
\author{Mitake {\sc Miyazaki},  Keita {\sc Kishigi} and
 Yasumasa {\sc Hasegawa}}
\inst{Faculty of Science, Himeji Institute of Technology,
\\Kamigouri-cho, Akou-gun, Hyogo 678-1297, Japan }
\recdate{}
\abst{We have investigated the transition temperature $T_{\rm c}(H)$ of superconductivity in quasi-two-dimensional (Q2D) tight-binding electrons in a strong magnetic field. When the magnetic field is parallel to 2D conducting plane, $T_{\rm c}(H)$ of the Q2D superconductor is shown to increase in an oscillatory manner as the magnetic field becomes large and to reach $T_{\rm c}(0)$ in a strong magnetic field limit for the spin-triplet superconductor. We consider the cases of on-site and nearest sites attractive interaction, and calculate the magnetic field dependences of the transition temperature for various types of symmetry. The first order transition from $p_y$-wave to $p_x$-wave is shown to occur at $H\sim 35$T when the magnetic field is parallel to the $y$ direction, which will be observed in a triplet superconductor, Sr$\mbox{}_2$RuO$\mbox{}_4$.}
\kword{field-induced superconductivity, organic conductors, 
quasi-two-dimension, Sr$\mbox{}_2$RuO$\mbox{}_4$}
\begin{document}
\sloppy
\maketitle

\section{introduction}

Recently, it has been shown that the quasi-two dimensional (Q2D) superconductivity evolves from the Ginzburg-Landau-Abrikosov-Gor'kov~\cite{rf:1,rf:2} (GLAG) and Lawrence-Doniach~\cite{rf:3,rf:4,rf:5} (LD) region to the reentrant phase
as a magnetic field becomes large~\cite{rf:6,rf:7}, which is derived by considering the quantum effect of electron motion in the presence of a strong 
magnetic field applied parallel to 2D conducting plane.
The Q2D systems with layered structure, for example high $T_{\rm c}$
oxides, $\beta$-(BEDT-TTF)$\mbox{}_2$I$\mbox{}_3$~\cite{rf:8} and Sr$\mbox{}_2$RuO$\mbox{}_4$,~\cite{rf:9} have a two-dimensional cylindrical Fermi surface with weak warping along the $k_z$ direction.  When the magnetic field $H$ is parallel to the
conducting plane, the electrons move in orbits which are the intersection of the Fermi surface and a plane normal to $H$ in a semi-classical picture. Some orbits are open and the others are closed. The former gives a similar effect as Q1D superconductivity in a strong magnetic field.~\cite{rf:10,rf:11,rf:12,rf:13,rf:14} By neglecting the contribution from the closed orbits, Lebed and Yamaji~\cite{rf:6} calculated the mean field transition temperature of a Q2D superconductor  in the parabolic band model and demonstrated the possibility of reentrant superconductivity. In a previous paper~\cite{rf:7}, we calculated the mean field transition temperature numerically by taking account of the eigenstates of three-dimensional tight-binding electrons in a strong magnetic field.
 In this formulation, we can treat Q1D and Q2D in the same manner by changing $t_b/t_a$, where $t_a$ and $t_b$ are hopping matrix elements in the conducting plane. In the previous calculation~\cite{rf:7}, however, we considered only on-site attractive interaction for the Q2D systems. 

The high $T_{\rm c}$ oxides are thought to be Q2D $d$-wave superconductors and $\kappa$-(BEDT-TTF$)_2$Cu[N(CN$)_2$]Br~\cite{rf:15,rf:16,rf:17,rf:18} is shown to be anisotropic Q2D with line nodes of the gap. Rice and Sigrist~\cite{rf:19} proposed that Sr$\mbox{}_2$RuO$\mbox{}_4$ could be a spin-triplet
superconductor by analogy with $\mbox{}^3$He and experimental evidences for the triplet superconductivity have been observed~\cite{rf:20,rf:21,rf:22}. The Q2D anisotropic superconductivity in a strong magnetic field is of great interest.

In this paper, we assume the nearest site attractive interaction in  tight-binding model, and we examine the transition temperature of anisotropic superconducting states in a strong magnetic field.

\section{Model and Green Function of Tight-Binding Electrons 
in a Magnetic Field}
The tight-binding electrons in a magnetic field are described by the Hamiltonian 
(we take $\hbar$, $k_{\rm B}$ and the velocity of light to be $1$), 
\begin{eqnarray}
{\cal H}&=&{\cal H}_0+{\cal H}_U\nonumber\\
{\cal H}_0&=&-t_a\sum_{(i,j)_a,\sigma}{\rm e}^{{\rm i}\theta_{ij}}
c^{\dag}_{i,\sigma}c_{j,\sigma}-t_b\sum_{(i,j)_b,\sigma}
{\rm e}^{{\rm i}\theta_{ij}}c^{\dag}_{i,\sigma}c_{j,\sigma}
\nonumber\\
& &-t_c\sum_{(i,j)_c,\sigma}{\rm e}^{{\rm i}\theta_{ij}}
c^{\dag}_{i,\sigma}
c_{j,\sigma}
-\mu\sum_{i,\sigma}c^{\dag}_{i,\sigma}c_{i,\sigma}\nonumber\\
& &-\sum_{i,\sigma}\sigma\mu_{\rm B}Hc^{\dag}_{i,\sigma}c_{i,\sigma}\\
{\cal H}_U&=&\sum_{<i,j>,\sigma,\sigma'}U_{ij}c^{\dag}_{i,\sigma}
c_{i,\sigma}c^{\dag}_{j,\sigma'}c_{j,\sigma'},
\end{eqnarray}
where $c^{\dag}_{i,\sigma}$ and $c_{i,\sigma}$ are creation and
annihilation operators at the $i$ site, $\mu$ is the chemical 
potential, $\sigma\mu_{\rm B}H$ is the Zeeman energy for $\uparrow$ 
($\downarrow$) spin ($\sigma=+(-)$), $U_{ij}$ is the interaction
between electrons at $i$ and $j$ sites and 
\begin{equation}
\theta_{ij}=\frac{2\pi}{\phi_0}\int^j_i\mbox{\boldmath
$A$}d\mbox{\boldmath$l$}.
\end{equation}

In the above, ${\mib A}$ is the vector potential and $\phi_0$ is the flux quantum. We consider the Q2D system with the hopping matrix elements $t_a\geq t_b\gg t_c$. The interlayer hopping matrix element $t_c$ is taken to be larger than transition temperature $T_{\rm c}$ for $H=0$, so that the fluctuation due to the low dimension is small and we take the mean field approximation in this paper.  

In the interaction Hamiltonian, eq.(2), we take the on-site interaction
and nearest-site interaction along each axis as
\begin{eqnarray}
U_{ij}=\left\{\begin{array}{ll}
U_0&\mbox{if}\ {\mib r}_i={\mib r}_j\\
U_\delta&\mbox{if}\ {\mib r}_i={\mib r}_j\pm \hat{{\mib \delta}}\\
0&\mbox{otherwise}\hspace{2cm}.
\end{array}
\right.
\end{eqnarray}
where $\hat{{\mib \delta}}=\hat{{\mib x}}$, $\hat{{\mib y}}$ and 
$\hat{{\mib z}}$ 
are unit vectors along $a$ axis, $b$ axis and $c$ axis, respectively.
The Fourier-transform of the interaction is obtained as
\begin{eqnarray}
U({\mib q})&=&U_0+2U_x\cos(aq_{x})+2U_y\cos(bq_{y})
\nonumber\\
& & +2U_z\cos(cq_{x}).
\end{eqnarray}

In this paper, the magnetic field $H$ is parallel to the $b$ axis. 
We use the vector potential ${\mib A}=(0,0,-Hx)$. Since the magnetic field considered in this paper is not extremely strong ($\phi/\phi_0\ll 1$, where $\phi=Hac$ is flux per unit cell), the change of the electron number as a function of $H$ may be very small in Q1D and Q2D case. This can be understood by noting that there are no closed orbits in the Q1D case and only small part of the orbits are closed in the Q2D case in the semi-classical picture. Therefore, the chemical potential $\mu$ is fixed for given $t_a$, $t_b$ and $t_c$ to give a quarter-filled band at $H=0$ instead of electron number to be fixed. 

The noninteracting Hamiltonian is written as 
\begin{eqnarray}
& &{\cal H}_0=\nonumber\\
& &\sum_{\sigma,\mbox{\scriptsize{\mib k}}}
C^{\dag}_{\sigma}
\left(\begin{array}{ccc}\ddots&&V^*\\&\begin{array}
{ccc}M_{-1}&V&0\\V^*&M_0&
V\\0&V^*&M_1\end{array}&\\V&&\ddots\end{array}\right)C_{\sigma},
\end{eqnarray}
where 
\begin{eqnarray}
M_{n}&=&-2t_a\cos[a(k_x+nG)]-2t_b\cos(bk_y)\nonumber\\
& &-\sigma\mu_{\rm B}H-\mu,
\end{eqnarray}
\begin{equation}
V=-t_c{\rm e}^{{\rm i}ck_z},
\end{equation}
\begin{equation}
C^{\dag}_{\sigma}=(\cdots,c^{\dag}_{\sigma}({\mib k}-{\mib
G}),c^{\dag}_{\sigma}
({\mib k}),c^{\dag}_{\sigma}({\mib k}+{\mib G}),\cdots),
\end{equation}
\begin{equation}
{\mib G}=(G,0,0)=\left(\frac{2\pi}{a}\frac{\phi}{\phi_0},0,0\right). 
\end{equation}

The size of the matrix in eq.(6) is $q\times q$ if $\phi/\phi_0=p/q$, where $p$ and $q$ are mutually prime integers and infinite if $\phi/\phi_0$ is irrational. The summation in ${\mib k}$ should be done in a magnetic Brillouin zone, $|k_x|<\pi/(qa)$, $|k_y|<\pi/b$, and $|k_z|<\pi/c$. The eigenvalues of the matrix in
eq.(6) for fixed $k_y$ have very rich structure as a function of the
magnetic field if $t_c\sim t_a$ as shown by Hofstadter~\cite{rf:24}. In the case of
$t_c\ll t_a$, which is the case for Q1D and Q2D systems, however, the
energy is almost continuous except for the large gaps near the bottom and top
of the energy. The gaps near the bottom and the top can be interpreted as the result of the Landau quantization in the
closed orbit and the negligible gaps can be understood as the small
probability of the magnetic breakdown in the open orbit. Since we
consider the instability to the superconductivity, only the states near
the Fermi surface with the energy $|\epsilon-\mu|<T$ are important.
For the electrons of less-than-half filled band, or the quarter-filled
electrons, we can consider only the part of the matrix in eq.(6) which
has, for example, $|a(k_x+nG)|<(3/4)(\pi/a)$.
By this approximation $\phi/\phi_0$ is not restricted to be a rational
number. The Brillouin zone should be taken as $|k_x|<G/2$ instead of the
magnetic Brillouin zone $|k_x|<\pi/(qa)=G/(2p)$ for the rational
magnetic field case $(\phi/\phi_0=p/q)$. The energy does not depend on
$k_z$ in this approximation. This can be seen by changing
$c^{\dag}_{\sigma}({\mib k}+m{\mib G})$ to ${\rm e}^{-{\rm
i}mck_z}c^{\dag}_{\sigma}({\mib k}+m{\mib G})$, by which only $(1,q)$ and $(q,1)$ elements in eq.(6) 
depend on $k_z$ but they are irrelevant for the energy near the Fermi
level in the case of $t_c\ll t_a$. The effects of the closed orbits and the magnetic breakdown are taken into account correctly in this approximation as shown in Fig. 1. 

We get the energy 
\begin{equation}
\varepsilon_{n,{\mib k},
\sigma}=\epsilon(n,k_x)-2t_b\cos(bk_y)
-\sigma\mu_{\rm B}H-\mu  
\end{equation}
and the eigenstates $|\Psi_{\sigma}(n,{\mib k})\rangle$ by numerically diagonalizing the matrix of size of
the order of $[(3/4)\cdot 2\pi/G]\times[(3/4)\cdot 2\pi/G]$.
\vspace{1.5cm}
\begin{figure}
\begin{center}
\leavevmode
\epsfxsize=9.5cm
\epsfbox{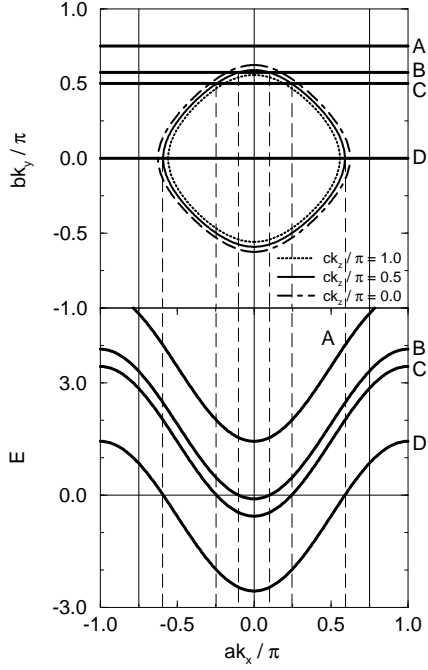}
\caption{Fermi surface in $k_x$-$k_y$ plane and energy as a function of $k_x$ in Q2D. Three curves in the upper figure are the Fermi surface in the planes of $k_z=\pi/c$, $\pi/(2c)$ and $0$, respectively. The lower figure shows the energy as a function of $k_x$ for four values of $k_y$ shown in the upper figure, representing no-orbit (A), closed orbit (B) and two open orbits (C) and (D) from the top. We neglect the region
of $|ak_x/\pi|>3/4$. In the presence of magnetic field gaps open at
$k_x=mG/2$ with integer $m\neq 0$.}
\label{fig1}
\end{center}
\end{figure}

The creation operators
$c^{\dag}_{\sigma}({\mib k}+m{\mib G})$ is expressed by the creation operators of the
eigenstates ($\Psi^{\dag}_{\sigma}(n,k)$) as 
\begin{equation}
c^{\dag}_{\sigma}({\mib k}+m{\mib G})={\rm e}^{{\rm i}mck_z}\sum_n
\phi^*_{k_x}(m,n)\Psi^{\dag}_{\sigma}(n,{\mib k}),
\end{equation}
where the coefficient $\phi_{k_x}(m,n)$ can be calculated numerically. Notice that $\phi_{k_x}(m,n)$ does not depend on $k_y$ and $k_z$.

The real space Green's function is given by 
\begin{eqnarray}
G_{\sigma}({\mib r}_i,{\mib r}_j',{\rm i}\omega_l)&=&-\int^{\beta}_0
{\rm d}\tau
{\rm e}^{{\rm i}\omega_l\tau}\left<T_{\tau}C_{{\mib r}_i,\sigma}(\tau)
C^{\dag}_{{\mib r}_j',\sigma}(0)\right>\nonumber\\
&=&\sum_{{\mib k},n}\sum_{m,m'}\frac{\phi_{k_x}
(m,n)\phi^*_{k_x}(m',n)}
{{\rm i}\omega_l-\varepsilon_{n,{\mib k},\sigma}}\nonumber\\
&\times& {\rm e}^{{\rm i}({\mib r}_j'
-{\mib r}_i)\cdot
{\mib k}+{\rm i}(m'{\mib r}_j'-m{\mib r}_i)\cdot{\mib G}}\nonumber\\
&\times& {\rm e}^{{\rm i}(m'-m)ck_z},
\end{eqnarray} 
where $\omega_l=(2l+1)\pi T$ is the Matsubara frequency.

The linearized gap equation  
for isotropic and anisotropic pairing in coordinate 
representation is obtained in the mean field approximation as 
\begin{equation}
\Delta_{\sigma\sigma'}({\mib r}_i,{\mib r}_j)=U_{ij}
\sum_s\sum_t
K_{\sigma\sigma'}({\mib r}_i,{\mib r}_j;{\mib r}_s',{\mib r}_t')
\Delta_{\sigma\sigma'}({\mib r}_s',{\mib r}_t'), 
\end{equation}
where
\begin{equation}
K_{\sigma\sigma'}({\mib r}_i,{\mib r}_j;{\mib r}_s',{\mib r}_t')
\equiv T\sum_{\omega_l}G_{\sigma}
({\mib r}_i,{\mib r}_s',{\rm i}\omega_l)G_{\sigma'}
({\mib r}_j,{\mib r}_t',-{\rm i}\omega_l).
\end{equation}

\section{Spin-triplet superconductivity}
In this section, we consider the spin-triplet superconductivity. We consider only the equal-spin-pairing state in the following.
The Zeeman effect does not play any important role in a strong magnetic field, as long as we neglect a change of the density of states of the up and down spin.

For the spin-triplet case, we define ``$p_{\delta}$-wave" the order parameter as
\begin{equation}
\Delta_{\uparrow\uparrow}^{p_{\delta}}({\mib r})\equiv
\frac{1}{2}\left[\Delta_{\uparrow\uparrow}({\mib r},{\mib r}+\hat{\mib \delta})-\Delta_{\uparrow\uparrow}({\mib r}+\hat{\mib \delta},{\mib r})\right].
\end{equation}
Here, the Fourier transformation of the order parameters is defined by 
\begin{equation}
\Delta_{N}^{p_\delta}\mbox{}_{\uparrow\uparrow}({\mib q})=\int d{\mib r} \Delta^{p_\delta}_{\uparrow\uparrow}({\mib r})
{\rm e}^{-{\rm i}({\mib q}+N{\mib G})\cdot ({\mib r}+\hat{\mib \delta}/2)},
\end{equation}
and 
\begin{equation}
\Delta^{p_\delta}_{\uparrow\uparrow}({\mib r})
=\sum_{{\mib q},N}
{\rm e}^{{\rm i}({\mib q}+N{\mib G})\cdot ({\mib r}+\hat{\mib \delta}/2)}
\Delta_{N}^{p_\delta}\mbox{}_{\uparrow\uparrow}({\mib q}),
\end{equation}
where $q_x$ is taken as $-G/2<q_x\le G/2$ and $N$ is integer. The 
$p_{\delta}$-wave state can be a solution of eq. (14) if $U_{\delta}<0$. When the system has tetragonal symmetry, $p_x$ and $p_y$ have the same coupling constant at $H=0$ and $p_x+{\rm i}p_y$ will be stabilized at $T<T_{\rm c}$, resulting in the nodeless gap in the Fermi surface. The magnetic field along the $b$ axis lifts the degeneracy of $p_x$ and $p_y$ even in a week magnetic field as discussed by Agterberg~\cite{rf:23} in the GL theory.

The linearized gap equation, eq.(14), is written as a matrix equation for $\Delta_{N}^{d_\delta}\mbox{}_{\uparrow\uparrow}({\mib q})$. The transition temperature is obtained by the condition that the nontrivial solution exists. It can be shown that the maximum transition temperature is obtained for ${\mib q}=0$. The matrix equation is reduced to the equations for even $N$ and odd $N$. For even $N$ the linearized gap equation is obtained as 
\begin{eqnarray}
\Delta_{2l}^{\cal T}\mbox{}_{\uparrow\uparrow}&=&U_{\delta}T\sum_{k_xk_y}\sum_{l'nn'}\sum_{\omega_l}\frac{1}{{\rm i}\omega_l-\varepsilon_{n,k_x,k_y\uparrow}}\frac{-1}{{\rm i}\omega_l+\varepsilon_{n',-k_x,-k_y\uparrow}}\nonumber\\
&\times&\sum_{m_1m_2}\sum_{m_3m_4}\delta_{m_1-m_2,m_3-m_4}\delta_{m_1+m_3,-N}\delta_{m_4-m_3,\frac{N}{2}-\frac{N'}{2}}\nonumber\\
&\times&\phi_{k_x}(m_1,n)\phi_{k_x}(m_2,n)\phi_{-k_x}(m_3,n')\phi_{-k_x}(m_4,n')\nonumber\\
&\times&\left[\cos\left\{\left(m_4-m_3-\frac{N}{2}+\frac{N'}{2}\right){\mib G}\cdot\hat{\mib \delta}\right\}\right.\nonumber\\
&-&\left.\cos\left\{2k_{\delta}+\left(m_4+m_3+\frac{N}{2}+\frac{N'}{2}\right){\mib G}\cdot\hat{\mib \delta}\right\}\right]\Delta_{2l'}^{\cal T}\mbox{}_{\uparrow\uparrow}\nonumber\\
&=&U_{\delta}\sum_{l'}\Pi_{2l,2l'}^{\cal T}\Delta_{2l'}^{\cal T}\mbox{}_{\uparrow\uparrow},
\end{eqnarray}
where 
\begin{eqnarray}
& &\Pi_{2l,2l'}^{\cal T}(q_x)=\sum_{k_x,k_y}\sum_{n,n'}\sum_m
(\gamma_{\mib k m}^{\cal T})^2
\nonumber\\
& &\times \phi_{k_x}(m-l,n)\phi_{k_x}(m-l',n)
\nonumber\\
& &\times \phi_{-k_x}(-m-l,n')\phi_{-k_x}(-m-l',n')
\nonumber\\
& &\times \frac{1-f(\varepsilon_{n,k_x,k_y\uparrow})
-f(\varepsilon_{n',-k_x-q_x,-k_y\uparrow})}
{2(\varepsilon_{n,k_x,k_y\uparrow}+\varepsilon_{n',-k_x-q_x,-k_y\uparrow})},
\end{eqnarray}
where $f(\varepsilon)$ is the Fermi distribution function and 
$\gamma_{\mib k m}^{\cal T}$ has the following forms for each order 
parameter:
\begin{eqnarray}
\gamma_{\mib k m}^{\cal T}=\left\{\begin{array}{cl}
\sin[a(k_x-mG)]&\mbox{${\cal T}$=$p_x$-wave}\\
\sin(bk_y)&\mbox{${\cal T}$=$p_y$-wave}\\
\sin(ck_z)&\mbox{${\cal T}$=$p_z$-wave}\hspace{2cm}.
\end{array}
\right.
\end{eqnarray}

For odd $N$, we get
\begin{equation}
\Delta_{2l+1}^{\cal T}\mbox{}_{\uparrow\uparrow}=U_{\delta}\sum_{l'}\Pi_{2l+1,2l'+1}^{\cal T}\Delta_{2l'+1}^{\cal T}\mbox{}_{\uparrow\uparrow},
\end{equation}
where
\begin{eqnarray}
& &\Pi_{2l+1,2l'+1}^{\cal T}(q_x)=\sum_{k_x,k_y}\sum_{n,n'}
\sum_m(\gamma_{\mib k m}^{\cal T})^2\nonumber\\
& &\times \phi_{k_x}(m-l,n)\phi_{k_x}(m-l',n)
\nonumber\\
& &\times \phi_{-k_x}(-m-l-1,n')\phi_{-k_x}(-m-l'-1,n')
\nonumber\\
& &\times \frac{1-f(\varepsilon_{n,k_x,k_y})
-f(\varepsilon_{n',-k_x-q_x,-k_y})}{2(\varepsilon_{n,k_x,k_y}
+\varepsilon_{n',-k_x-q_x,-k_y})}.
\end{eqnarray}

The transition line is given by $1-gU_{\delta}=0$, where $g$ is the maximum eigenvalue of the matrix $\Pi$ for even and odd $N$. In this paper, we calculate the field dependence of $g$ at low temperature instead of calculating the transition temperature. 

In the following, we consider the Q2D superconductor of the quarter-filled electrons with parameters $t_b/t_a=1$, $t_c/t_a=0.05$ and $T/t_a=0.001$. In Fig. 2, we plot the effective coupling constant $g/g_0$ for each states as a function of $aG/(2\pi)=\phi/\phi_0$, where $g_0$ is the effective coupling constant for $t_c=0$, which corresponds to that in the absence of magnetic field. For example, the magnetic field of $20$ Tesla corresponds to $\phi/\phi_0\simeq 0.0036$ for $a=3.87\AA$ and $c=12.7\AA$ (Sr$\mbox{}_2$RuO$\mbox{}_4$). It is found that $g/g_0$ obtained by diagonalizing the even part of the matrix $\Pi$ increases in an oscillatory manner as a magnetic field becomes larger and reaches that in the absence of magnetic field, while $g/g_0$ for the odd part becomes zero in the strong field limit. In the semi-classical treatment of the magnetic field, superconductivity is destroyed in a strong magnetic field, since the energies of electrons with wave numbers $\mib k$ and $-\mib k$ is not equal. However, as is seen in eq. (19), the instability to the superconductivity is caused by the pairs between the states ($n,\mib k$) and ($n',-\mib k$) with the same energy $\varepsilon_{n,k_x,k_y}=\varepsilon_{n',-k_x,-k_y}$. On
the other hands, the coefficient $\phi_{k_x}(m,n)$ becomes small in a weak magnetic field region but a lot of states within the energy
range of $T$ contribute to forming Cooper pairs, which will reproduce the GLAG result.

\begin{figure}
 \begin{center}
 \leavevmode
 \epsfxsize=7.5cm
 \epsfbox{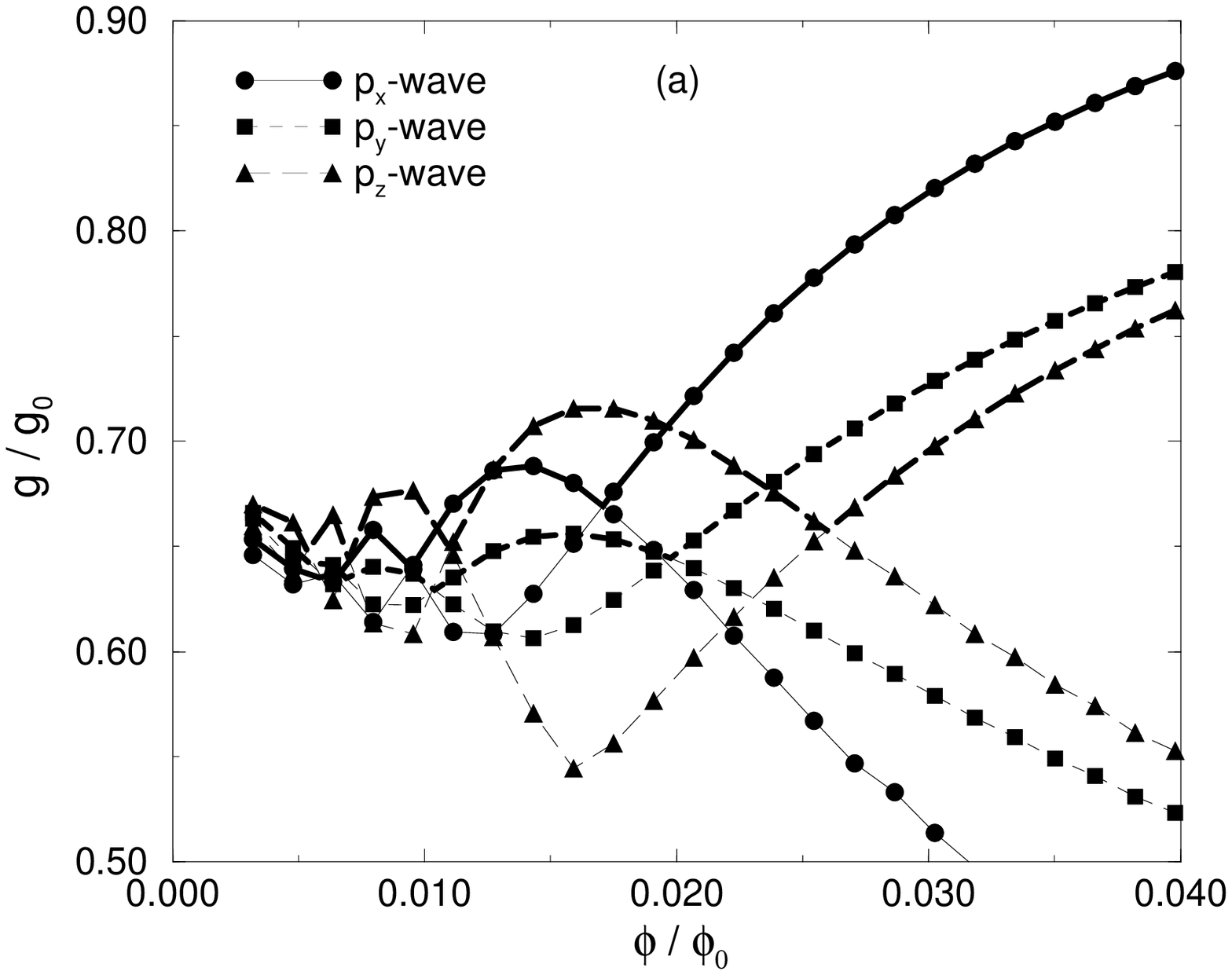}
\vspace{7mm}
 \leavevmode
 \epsfxsize=7.5cm
 \epsfbox{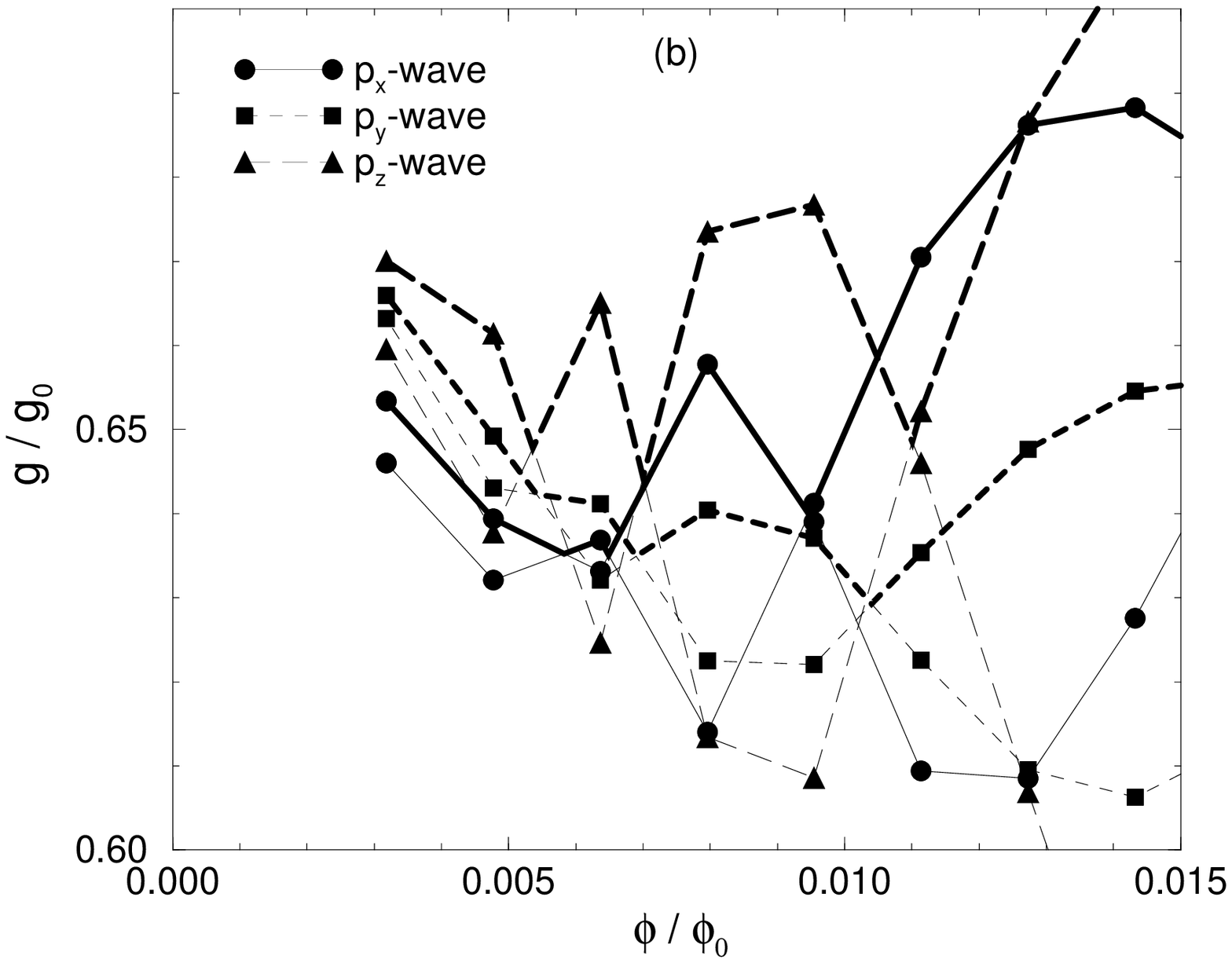}
\caption{(a) Effective coupling constant as a function of $\phi/\phi_0$
 in the case of $t_b/t_a=1.0$, $t_c/t_a=0.05$, $T/t_a=0.001$ and ${ \mib H}\parallel y$. Two lines for each state are obtained from the even and odd parts. The larger $g/g_0$ are shown by thick lines. (b) The low field region.}
\label{fig2}
 \end{center}
\end{figure}

We obtain $g/g_0$ is larger in the $p_x$-wave state than in the $p_y$-wave state for $\phi/\phi_0\geq 0.0065$ as shown in Fig. 2. The difference of $g/g_0$ can be understood as follows. Since the electron orbit at $k_y\approx0$ is open, we can use the linearized dispersion and we can treat it as a quasi-one dimensional system~\cite{rf:7}. Then we obtain $\phi_{k_x}(m,n)\approx J_{m-n}\left(2t_c/v_{\rm F}(k_y)G\right)$, where $J$ is the Bessel function and $v_{\rm F}(k_y)$ is the $k_y$-dependent $k_x$ component of the Fermi velocity~\cite{rf:7}. Therefore, for the larger $v_{\rm F}(k_y)$, the larger effect of the magnetic field is expected. On the other hand, the electron orbit at $k_y\approx\pi/2b$ in the quarter filled band is closed or has a small $v_{\rm F}(k_y)$ even if the orbit is open (see Fig. 1). In the $p_x$-wave state the amplitude of the order parameter is largest at $k_x=\pi/2a$, which corresponds to $k_y\approx 0$ in the Fermi surface of the quarter filled band. The order parameter in $p_y$-wave state is zero at $k_y=0$. Thus the quantum effect, which causes the increase of $g/g_0$ in a strong magnetic field, is larger in the $p_x$-wave state than that in the $p_y$-wave state. This is a simple interpretation using the Fermi surface in $H=0$. In the case of $H\neq 0$, we can interpret in the same way as follows. By using the approximation of the linearized dispersion, we can write eq. (20) for the $p_x$ and $p_y$ wave as,
\begin{eqnarray}
\Pi_{2l,2l'}^{\cal T}&=&\sum_{k_x,k_y}\left\{\begin{array}{c}
\sin^2a(k_x-k_{\rm F}(k_y))\\\sin^2bk_y\\ \end{array}\right\}\sum_N\nonumber\\
&\times&\frac{1-f(v_{\rm F}(k_y)(k_x-k_{\rm F}(k_y)))
-f(v_{\rm F}(k_y)(k_x+NG-k_{\rm F}(k_y)))}
{v_{\rm F}(k_y)(2k_x+NG-2k_{\rm F}(k_y))}\nonumber\\
&\times&\int d\theta J_{N-2l}\left(\frac{2t_c}{v_{\rm F}(k_y)G}\sin\theta\right)J_{N-2l'}\left(\frac{2t_c}{v_{\rm F}(k_y)G}\sin\theta\right),
\end{eqnarray}
where $k_{\rm F}(k_y)$ is the $k_y$-dependent Fermi wave number. The large contribution comes from the region of large $v_{\rm F}(k_y)$, i.e. $k_y\approx 0$, where $\sin^2a(k_x-k_{\rm F}(k_y))>\sin^2bk_y$. As a result, the $g/g_0$ for $p_x$-wave is larger than that for $p_y$-wave.
In the tetragonal system with $t_a=t_b\gg t_c$ and $U_x=U_y\gg U_z$, the $p_x$ state is realized at the transition line in the strong magnetic field along the $b$ axis ($y$ direction), and the $p_y$ component will become finite at the second transition at lower temperature. 

In the weak magnetic field, we can apply the GL theory. As shown by Agterberg~\cite{rf:23}, only the $p_y$-wave orders at $H_{\rm c2}$ (or at the transition temperature $T_{\rm c}(H)$) when $H$ is along the $b$ axis, and at the second transition the $p_x$-wave becomes finite. The difference of the transitions for the $p_x$- and $p_y$-waves comes from the difference of the GL parameters $\kappa_1$ and $\kappa_2$ ($\kappa_1>\kappa_2$), which are the coefficients of the gradient terms ($|D_x\Delta^{p_x}_{\uparrow\uparrow}({\mib r})|^2+|D_y\Delta^{p_y}_{\uparrow\uparrow}({\mib r})|^2$) and ($|D_y\Delta^{p_x}_{\uparrow\uparrow}({\mib r})|^2+|D_x\Delta^{p_y}_{\uparrow\uparrow}({\mib r})|^2$), respectively, where $D_i=\nabla_i-eA_i$. 

At a first glance, the GL theory and our result obtained in the strong field may be considered to be incompatible, but it is reasonable. As discussed above the quantum effect of the magnetic field to increase $g/g_0$ in a strong magnetic field is stronger in the $p_x$-wave than in the $p_y$-wave for $H\parallel b$. In the same way the orbital frustration effect to reduce the transition temperature in a weak magnetic field is stronger in the $p_x$-wave than in the $p_y$-wave, resulting in the lower transition temperature in the $p_x$-wave in a weak magnetic field. In the crossover region of the magnetic field relative value of $g/g_0$ for the $p_x$-wave and the $p_y$-wave will change. In Fig. 2(b), we see these features at $\phi/\phi_0\approx 0.0065$ ($H\sim 35$T).

\section{Spin-singlet superconductivity}
In this section, we consider the spin-singlet superconductivity. 
In the presence of the nearest-site interaction, $\Delta_{\uparrow\downarrow}({\mib r},{\mib r}+\hat{\mib \delta})=
-\Delta_{\downarrow\uparrow}({\mib r},{\mib r}+\hat{\mib \delta})=\Delta_{\uparrow\downarrow}({\mib r}+\hat{\mib \delta},{\mib r})=-\Delta_{\downarrow\uparrow}({\mib r}+\hat{\mib \delta},{\mib r})$. Therefore, the order parameters of ``$d_\delta$-wave" is defined by
\begin{equation}
\Delta^{d_\delta}_{\uparrow\downarrow}({\mib r})\equiv\frac{1}{2}
[\Delta_{\uparrow\downarrow}({\mib r},{\mib r}+\hat{\mib \delta})+\Delta_{\uparrow\downarrow}({\mib r}+\hat{\mib \delta},{\mib r})]
\end{equation}
The $d_x$-wave or $d_y$-wave is realized only if $U_x<0$ and $U_0=0$ or $U_y<0$ and $U_0=0$, respectively. In contrast to the spin-triplet superconductivity, if $U_0\neq 0$, $d_x$ and $d_y$ are not self-consistent solutions of eq. (14) but the linear combination of $\Delta^{d_x}_{\uparrow\downarrow}(r)$, $\Delta^{d_y}_{\uparrow\downarrow}(r)$ and $\Delta^{s}_{\uparrow\downarrow}\equiv\Delta_{\uparrow\downarrow}(r,r)$ can be a self-consistent solution. If the system has a tetragonal symmetry ($t_a=t_b$ and $U_x=U_y<0$), the $d$-wave ($d_x-d_y$) is realized at $H=0$. However, $d_x-d_y$ is not the self-consistent solution in the presence of the magnetic field along the $b$ axis. The $d_z$ is realized as long as $U_z<0$ and the higher terms in $t_c/t_a$ are neglected.
In this section we consider the case that one of the interactions ($U_0$ or $U_{\delta}$) is negative and the others are zero, so that $s$ or $d_{\delta}$-wave is realized, for simplicity. 
Here, we write
\begin{equation}
\Delta^{d_\delta}_{\uparrow\downarrow}({\mib r})=\sum_{{\mib q},N}{\rm e}^{{\rm i}({\mib q}+N{\mib G})\cdot ({\mib r}_i+\hat{\mib \delta}/2)}\Delta^{d_{\delta}}_{N}\mbox{}_{\uparrow\downarrow}({\mib q}).
\end{equation}
By taking $\hat{\mib \delta}=0$, we get the $s$-wave order parameter.

The linearized gap equation is written as a matrix equation for 
$\Delta^{d_{\delta}}_{N}\mbox{}_{\uparrow\downarrow}({\mib q})$ as in the spin-triplet case. The matrix equation is separated into even $N$ and odd $N$ parts. We get
\begin{equation}
\Delta_{2l}^{\cal S}\mbox{}_{\uparrow\downarrow}=\lambda\sum_{l'}\Pi_{2l,2l'}^{\cal S}\mbox{}_{\uparrow\downarrow}
\Delta_{2l'}^{\cal S}\mbox{}_{\uparrow\downarrow},
\end{equation}
where coupling constant $\lambda$ is $U_0$ or $U_\delta$ and  
\begin{eqnarray}
& &\Pi_{2l,2l'}^{\cal S}=\sum_{k_x,k_y}\sum_{n,n'}\sum_m
(\gamma_{\mib k m}^{\cal S})^2
\nonumber\\
& &\times \phi_{k_x}(m-l,n)\phi_{k_x}(m-l',n)
\nonumber\\
& &\times \phi_{-k_x}(-m-l,n')\phi_{-k_x}(-m-l',n')
\nonumber\\
& &\times \frac{1-f(\varepsilon_{n,k_x,k_y,\uparrow})
-f(\varepsilon_{n',-k_x,-k_y,\downarrow})}
{2(\varepsilon_{n,k_x,k_y,\uparrow}+
\varepsilon_{n',-k_x,-k_y,\downarrow})},
\end{eqnarray}
where $\gamma_{\mib k m}^{\cal S}$ has the following forms 
for each order parameter:
\begin{eqnarray}
\gamma_{\mib k m}^{\cal S}=\left\{\begin{array}{cl}
1&{\cal S}\mbox{=$s$-wave}\\
\cos[a(k_x-mG)]&{\cal S}\mbox{=$d_x$-wave}\\
\cos(bk_y)&{\cal S}\mbox{=$d_y$-wave}\\
\cos(ck_z)&{\cal S}\mbox{=$d_z$-wave}\hspace{2cm}.
\end{array}
\right.
\end{eqnarray}

For odd $N$, we get
\begin{equation}
\Delta_{2l+1}^{\cal S}\mbox{}_{\uparrow\downarrow}=\lambda\sum_{l'}\Pi_{2l+1,2l'+1}^{\cal S}\mbox{}_{\uparrow\downarrow}
\Delta_{2l'+1}^{\cal S}\mbox{}_{\uparrow\downarrow},
\end{equation}
where
\begin{eqnarray}
& &\Pi_{2l+1,2l'+1}^{\cal S}=\sum_{k_x,k_y}\sum_{n,n'}
\sum_m(\gamma_{\mib k m}^{\cal S})^2\nonumber\\
& &\times \phi_{k_x}(m-l,n)\phi_{k_x}(m-l',n)
\nonumber\\
& &\times \phi_{-k_x}(-m-l-1,n')\phi_{-k_x}(-m-l'-1,n')
\nonumber\\
& &\times \frac{1-f(\varepsilon_{n,k_x,k_y,\uparrow})
-f(\varepsilon_{n',-k_x,-k_y,\downarrow})}
{2(\varepsilon_{n,k_x,k_y,\uparrow}
+\varepsilon_{n',-k_x,-k_y,\downarrow})}.
\end{eqnarray}

In the following, we neglect the Zeeman energy for simplicity to show the different behavior between each state obviously.
We first calculate the effective coupling constant of $s$-wave superconductivity in a strong magnetic field and examine its $t_c/t_a$ dependence.

\begin{figure}
 \begin{center}
 \leavevmode
 \epsfxsize=7.5cm
 \epsfbox{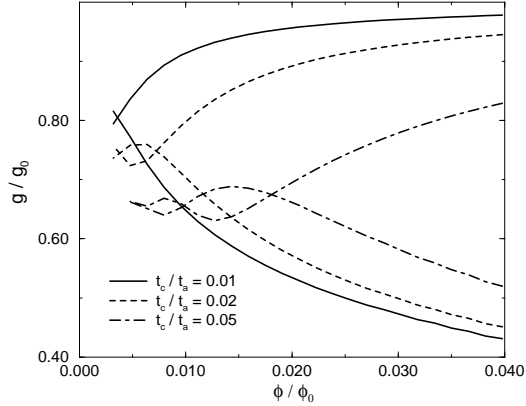}
\caption{Effective coupling constant of the s-wave pairing as a function of $\phi/\phi_0$ in the case of $t_b/t_a=1.0$, $T/t_a=0.001$ and ${\mib H}\parallel y$.}
\label{fig3}
 \end{center}
\end{figure}

In Fig. 3, we plot the effective coupling constant as a function of 
$\phi/\phi_0$. The effective coupling constant reaches that for $t_c=0$ as a magnetic field increases. We find that the effective coupling constant depends strongly on the hopping matrix elements between layers $t_c$. As $t_c/t_a$ becomes small, the oscillation of $g/g_0$ becomes small, and the value of $g/g_0$ increases in whole. Thus, reentrant behavior will be observed in weaker magnetic field in the superconductor with smaller $t_c/t_a$.

We now study the effective coupling constant of Q2D anisotropic superconductor with $t_c/t_a=0.05$. In Fig. 4, we plot $g/g_0$ obtained by each pairing state as a function of $\phi/\phi_0$.

\begin{figure}
 \begin{center}
 \leavevmode
 \epsfxsize=7.5cm
 \epsfbox{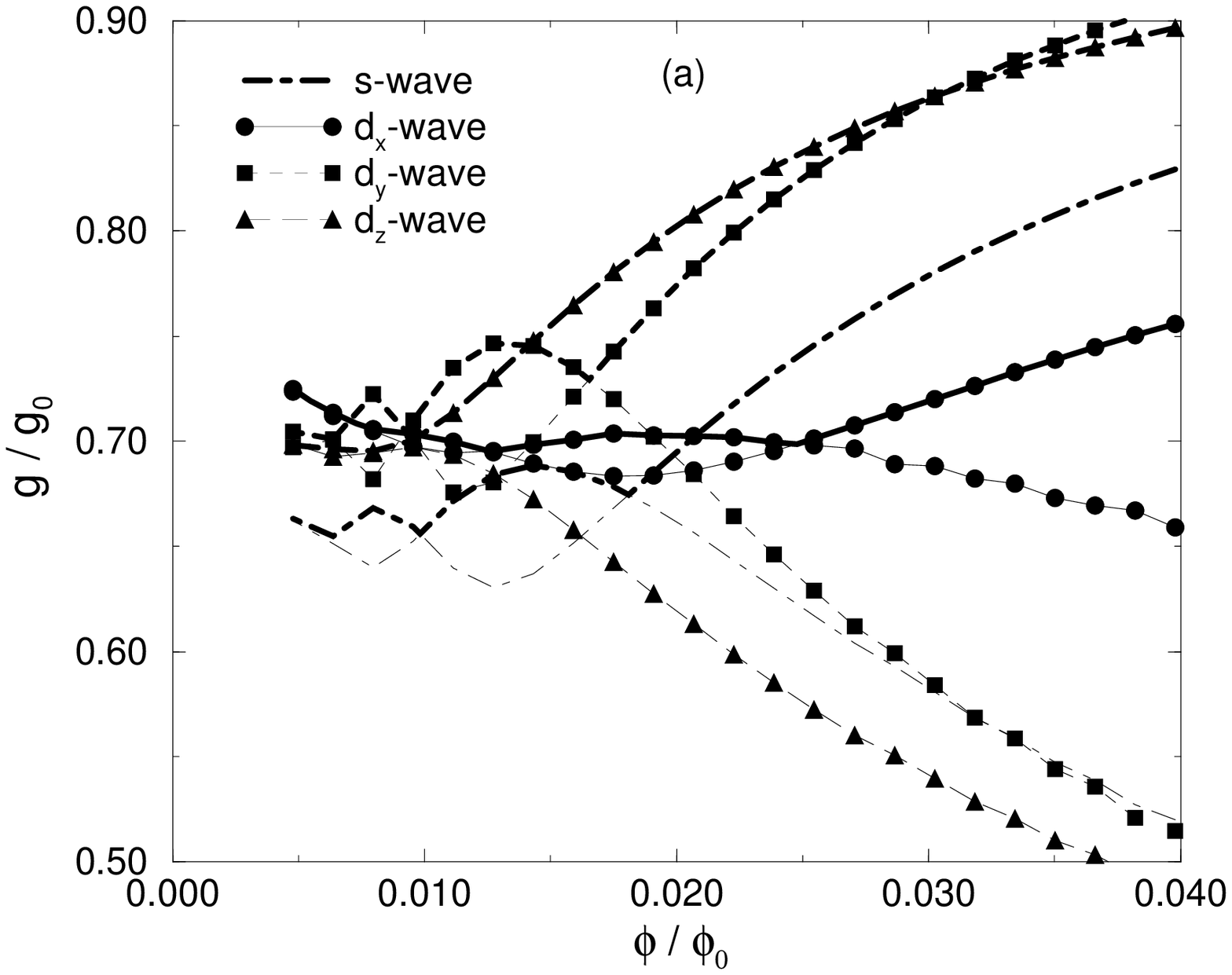}
\vspace{7mm}
 \leavevmode
 \epsfxsize=7.5cm
 \epsfbox{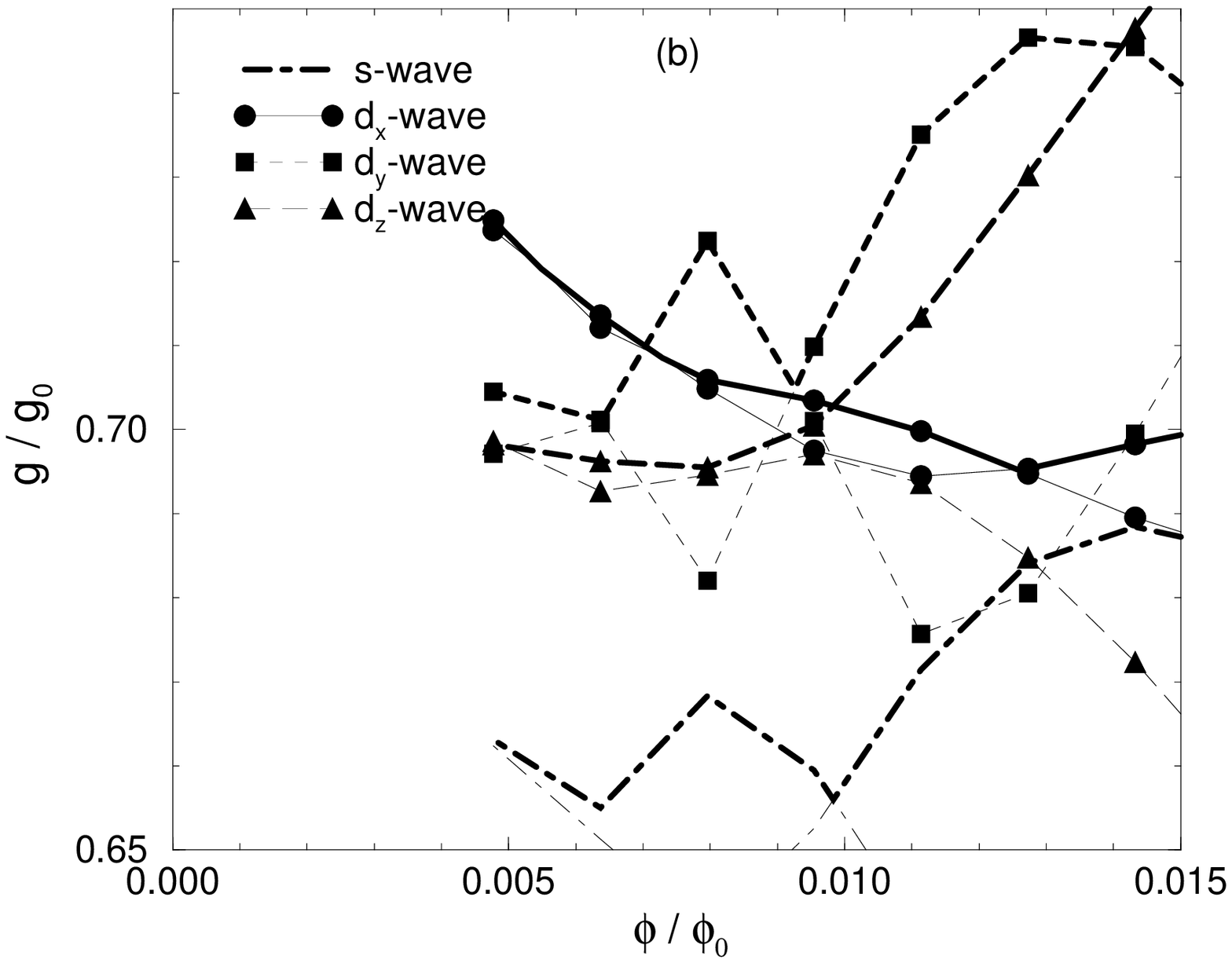}
\caption{(a) Effective coupling constant as a function of $\phi/\phi_0$
 in the case of $t_b/t_a=1.0$, $t_c/t_a=0.05$, $T/t_a=0.001$ and ${\mib H}\parallel y$. Two lines for each state are obtained from the even and odd parts. The larger $g/g_0$ are shown by thick lines. (b) The low field region.}
\label{fig4}
 \end{center}
\end{figure}

For each order parameter, $g/g_0$ of the even part increases and that of the odd part decreases in a strong magnetic field limit. The effective coupling constant $g/g_0$ in the $d_y$-wave state is larger than that in the $d_x$-wave state. This can be understood as in the spin-triplet case. The $d_y$-wave have a largest order parameter at $k_y=0$ and its order parameter is zero at $k_y=\pi/2b$, while the order parameter of the $d_x$-wave is largest at $k_x=0$ ($k_y\approx \pi/2b$ for the quarter filled band) and zero at $k_x=\pi/2a$ ($k_y\approx 0$). As a result, the recovery of $g/g_0$ in the strong magnetic field due to the quantum effect in the open orbit occurs in lower magnetic field for the $d_y$ state than for the $d_x$ state as seen in Fig. 4(b). 

\section{Conclusion}
In this paper we study the anisotropic superconductivity of
tight-binding electrons with hopping matrix elements $t_a=t_b\gg
t_c$ as a function of the magnetic field parallel to the $b$ axis. We
calculate the energies and the eigenstates in the magnetic field by
numerically diagonalizing the matrix and the effective coupling constant is calculated by using these values. The effects of both open and
closed orbits are taken into account in our calculation. We consider
the attractive interaction between electrons in the nearest sites along
each axis to realize the anisotropic superconductivity. With these interactions singlet ($d_x$, $d_y$ and $d_z$) and
triplet ($p_x$, $p_y$ and $p_z$) superconductivities are possible. 
The effective coupling constant, $g/g_0$, for these anisotropic
superconductivities are calculated. It is obtained that $g/g_0$
approaches to 1 as $H$ becomes large and it depends on the symmetry of
the order parameter. As shown in Figs. 2 and 4, we find that the $p_x$-wave ($d_y$-wave) state gives higher transition temperature than that of $p_y$-wave ($d_x$-wave) in the strong magnetic field region, and  both of them reach the transition temperature of zero magnetic field in a strong field limit. The first order transition from $p_y$-wave to $p_x$-wave is predicted at $H\sim 35$T, which will be observed in the $p$-wave superconductor, Sr$\mbox{}_2$RuO$\mbox{}_2$.

\section{Acknowledgment}
One of the authors (Y. H) thank D. F. Agterberg and Z. Q. Mao for useful discussions. This work was partially supported by Grant-in-Aid for JSPS Fellows from the Ministry of Education, Science, Sports and Culture. One of the authors (K. K) was financially supported by the Research Fellowships of the Japan Society for the Promotion of Science for Young Scientists.

\end{document}